\documentclass{Interspeech2024}
\interspeechcameraready 

\title{TSE-PI: Target Sound Extraction under Reverberant Environments with Pitch Information}

\name[affiliation={1}]{Yiwen}{Wang}
\name[affiliation={1,2,3}]{Xihong}{Wu}

\address{
  $^1$Speech and Hearing Research Center, School of Intelligence Science and Technology, Peking University, Beijing, China 
  $^2$National Key Laboratory of General Artificial Intelligence\\
  $^3$Institute for Artificial Intelligence, Peking University
}
\email{pku\_wyw@pku.edu.cn, wxh@cis.pku.edu.cn}


\begin{document}

\maketitle

\begin{abstract}
Target sound extraction (TSE) separates the target sound from the mixture signals based on provided clues. However, the performance of existing models significantly degrades under reverberant conditions. Inspired by auditory scene analysis (ASA), this work proposes a TSE model provided with pitch information named TSE-PI. Conditional pitch extraction is achieved through the Feature-wise Linearly Modulated layer with the sound-class label. A modified Waveformer model combined with pitch information, employing a learnable Gammatone filterbank in place of the convolutional encoder, is used for target sound extraction. The inclusion of pitch information is aimed at improving the model's performance. The experimental results on the FSD50K dataset illustrate 2.4 dB improvements of target sound extraction under reverberant environments when incorporating pitch information and Gammatone filterbank.
\end{abstract}

\section{Introduction}
The cocktail party problem shows that humans have an extraordinary ability at the selective auditory attention to the target sound under complex acoustic environments, such as noise and reverberation \cite{cherry1953cocktail}. Motivated by the need to bridge the gap between human auditory perception and machine hearing, researchers have developed TSE models \cite{gfeller2021one, tzinis22_interspeech, delcroix2022soundbeam, ohishi2022conceptbeam, veluri2023real, veluri2023semantic, zmolikova2023neural, hai2024dpm, kim2024improving}. Target sound extraction aims to separate the desired sound from a mixture of various sound events, given a specific clue leading to the target sound event. Common clue conditions can be divided into classes \cite{veluri2023real, veluri2023semantic}, enrollment information \cite{gfeller2021one, delcroix2022soundbeam}, query-based separation \cite{chen2022zero, liu22w_interspeech}, etc. In addition, multi-cue and multi-modal cues are also used to achieve separation \cite{tzinis22_interspeech, li2023target, ye2024lavss, veluri2024look}.

Most of the above methods have achieved excellent performance under anechoic sound conditions. However, there remains a gap between the performance of the TSE models under complex environments and the human auditory system. Recently, there have been several discussions on target sound extraction under reverberation. These models aim to extract desired sounds from complex acoustic mixtures, such as those encountered in the real world. Veluri et al. proposed a real-time Waveformer for binaural processing, which can be applied to real scenarios \cite{veluri2023semantic}. Choi and Choi introduced a transformer-based TSE model to extract reverberant sounds using the Dense Frequency-Time Attentive Network (DeFT-AN) architecture \cite{lee2023deft, choi2023target}. The complex short-time Fourier transform (STFT) mask is generated by supplying the sound class label. These methods have specific effects under reverberation conditions but must still be closer to the results under anechoic conditions.

To further improve the target sound extraction performance under reverberant environments, it is necessary to refer to the robustness of the auditory system. ASA is a critical process for understanding and interpreting complex sound environments where multiple sound sources coexist \cite{bregman1994auditory}. Pitch information plays a vital role in the ASA process. Pitch, corresponding to the harmonics' fundamental frequency (f0), contributes to the perceptual segregation. The theory of computational auditory scene analysis (CASA), proposed by Wang and Brown \cite{wang2006computational}, shows that pitch information, regarded as a discriminative clue, is helpful for bottom-up foreground separation \cite{shamma2011temporal}.

Auditory systems are robust, whatever the complex acoustic scene is. Temporal coherence analysis shows that humans simultaneously tend to focus on a single auditory stream. In the conventional CASA, there is a process of top-down auditory selective attention and bottom-up auditory stream formation \cite{shamma2011temporal}. Tasks for target sound extraction are simplified. That is, the separation of foreground sounds is achieved on the premise that clues such as categories are provided. Under such an assumption, only the bottom-up foreground segregation is considered. During the bottom-up process, pitch information is the leading perceptual feature to be noticed. Therefore, referring to CASA, we propose a two-stage target sound extraction model in complex acoustic scenarios. Specifically, a conditional pitch extraction model is proposed to extract the target pitch information belonging to the target sound. With the pitch information, direct sound is separated with a modified Waveformer architecture. The main contributions of this paper are:

\begin{itemize}
     \item A two-stage target sound extraction network is proposed. For the first stage, pitch information of the target direct sound under reverberation conditions is extracted.
     %with the Feature-wise Liearly Modulated (FiLM) layer \cite{perez2018film}
     For the second stage, target sound extraction is achieved with pitch information extracted from the first stage. A modified Waveformer is chosen as the target sound separation network. \footnote{Code of our work is available on \url{https://github.com/wyw97/TSE_PI}}
     \item A learnable Gammatone filter bank is introduced for conditional sound source separation. The pitch information contains frequency information, and the Gammatone filterbank often simulates the spectral analysis of the cochlea \cite{lopez2001human}.
     \item The proposed target sound extraction model guided by pitch information brings about 2.4 dB improvements under reverberant conditions. Experimental results show that the bottom-up foreground sound separation in the CASA framework has essential guidance for the TSE task.
\end{itemize}

The subsequent sections of the paper are organized as follows: Section 2 introduces the proposed two-stage pitch-guided target sound extraction. The experimental setup is described in Section 3, while the experimental results are reported in Section 4. Finally, conclusions are drawn in Section 5.

\begin{figure*}[htbp]
    \centering
    \includegraphics[width=0.75\textwidth]{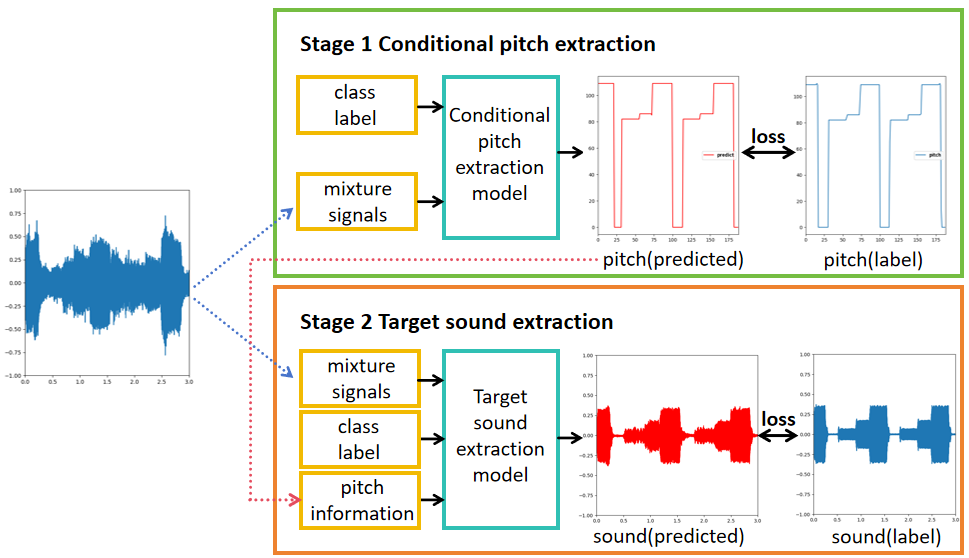}
    \caption{Overview of the proposed two-stage target sound extraction with pitch information (TSE-PI).}
    \label{fig:twostage}
\end{figure*}

\section{Method}

The pipeline of the proposed two-stage target sound extraction model with pitch information (TSE-PI) is shown in Figure \ref{fig:twostage}. This section discusses the implementation methods of these two stages in detail. In this study, a single-channel received signal $y_{M, L}\in R^{T}$ positioned at location $L$ comes from $N$ sound sources $s_{n}(n=1, ...,N)$. $T$ is the signal duration. The received mixture signal is given as 
\begin{equation}
    y_{M, L} =\sum_{i=1}^{N} s_{i} \ast h_{i, L} + bn,
\end{equation}
where $h_{i, L}$ represents the impulse response from the sound source to the receiver position $L$, $\ast$ represents convolution, and $bn$ refers to the background noise. For the given class label $c$, supposing that the sound source $s_{ic}$ belongs to class $c$,  the goal of the work is to separate from the mixture signal to obtain the direct-path signal of $s_{ic}$,
\begin{equation}
    \hat{y}_{ic, L} = s_{ic} \ast d_{ic, L},
\end{equation}
where $\hat{y}_{ic, L}$ denotes the direct-path signal generated from $s_{ic}$, and $d_{ic, L}$ is the direct part of the corresponding $h_{ic, L}$.

\subsection{Stage 1: Conditional pitch extraction}

Pitch information of the given class label is estimated through the first stage. The pitch feature is extracted from the harmonic structure of the signal amplitude spectrum. Recently, there have been several representative works using deep learning to achieve pitch extraction \cite{kim2018crepe, li2023pgss, singh2021deepf0, wei23b_interspeech}. Convolution models are commonly used to extract spectral features, and fully connected layers (FCL) are selected for mapping from harmonic features to pitch information. Current works perform well in multi-pitch and multi-track pitch extraction tasks. To enable the model to pay attention to the pitch information of a specific class of sounds, the Feature-wise Liearly Modulated (FiLM) layer is introduced to achieve target pitch extraction by modulating the output channel of each convolutional layer \cite{perez2018film}. We follow the basic framework for pitch estimation with a temporal convolutional network (TCN) as described in \cite{li2023pgss}. 

\begin{figure}[htbp]
    \centering
    \includegraphics[width=0.48\textwidth]{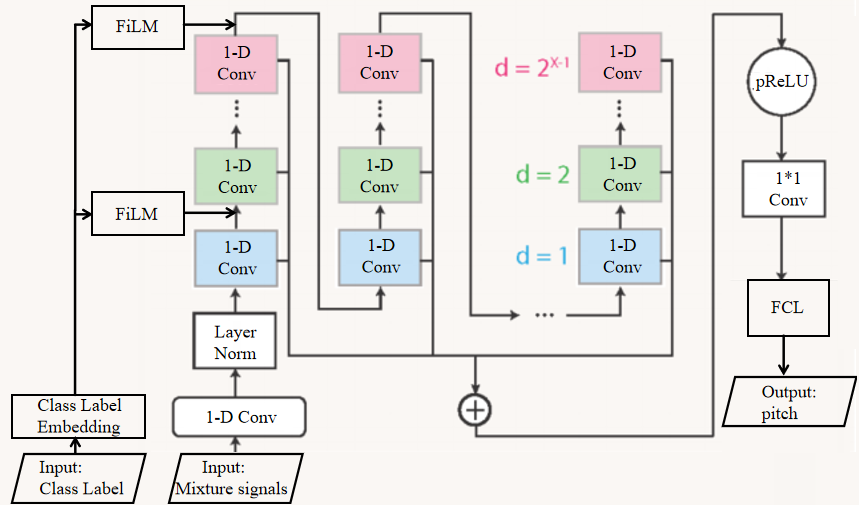}
    \caption{Condition pitch extraction model with FiLM. (To clearly display the FiLM module, only two FiLMs are drawn in the figure. In the actual model, each convolutional layer is modulated using FiLM.)}
    \label{fig:film_tcn}
\end{figure}

Specifically, for each output $F^{l}\in R^{c \times h \times w}$ from the $l$-th convolutional layer, where $c$ is the kernel number, $h$ and $w$ represents the width and height of $F^{l}$. The FiLM modulates each layer as 
\begin{equation}
    FiLM(F_{i}^{(l)}|\gamma_{i}^{(l)}, \beta_{i}^{(l)}) = \gamma_{i}^{(l)}F_{i}^{(l)} + \beta_{i}^{(l)},
\end{equation}
where $F_{i}^{(l)} \in R^{h \times w}$, $\gamma_{i}^{(l)}, \beta_{i}^{(l)} \in R^{c}$ refers to the corresponding modulation parameters. The modulation parameters are trained together with the other parameters of the model. The details of the other parts of the model are introduced in \cite{li2023pgss, luo2019conv}.

\subsection{Stage 2: Target sound extraction}
In the second stage, pitch information is added based on the existing target sound extraction model proposed by Veluri et al. \cite{veluri2023real}. Since pitch information plays a vital role in the spectral features, the pitch information extracted from the previous stage is concatenated with the features of the mixture signal along the channel dimension. Features are extracted through the 1-D convolutional encoder. The pitch information obtained in the first stage is expressed as a one-hot encoding, consistent with the method described in \cite{li2023pgss}.
% In this work, we focus on using pitch information cues to help sound source separation. Research on how to better implement pitch-guided sound source separation will be left in the following research. 
\begin{figure}[htbp]
    \centering
    \includegraphics[width=0.48\textwidth]{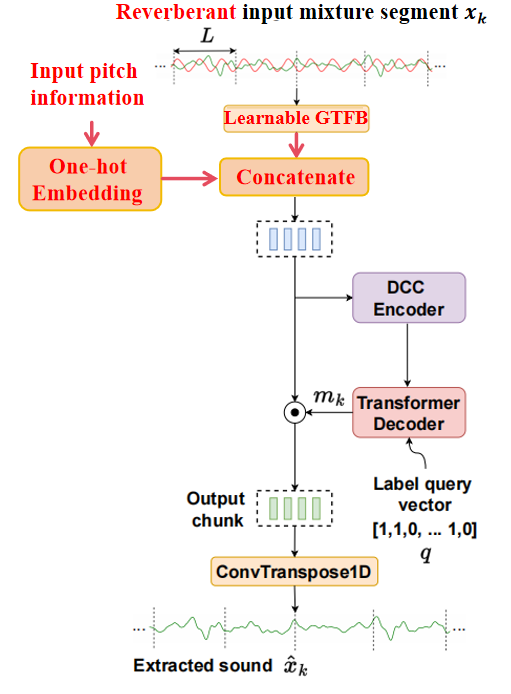}
    \caption{Target sound extraction with pitch information. (The part marked in \textcolor{red}{red} is the modification on Waveformer.)}
    \label{fig:enter-label}
\end{figure}

By adding pitch information, explicit spectral representation is more intuitive than the convolutional encoder. Inspired by the excellent performance of the auditory system, Gammatone filterbank (GTFB) is introduced in spectral analysis to further increase the connection to pitch information. GTFB with learnable parameters benefits universal sound separation (USS) performance, as shown in \cite{li2021auditory}. Referring to this work, we apply the encoding process based on the learnable GTFB to the target sound extraction under reverberant conditions. 

\section{Experimental Framework}
\subsection{Datasets introduction}
Experiments are carried out on FSD50K datasets \cite{fonseca2021fsd50k}. Twenty-seven sound classes are selected from the datasets, as shown in Figure \ref{fig:cmp_sisnr_plot}. Each class contains at least 40 4-second samples. Reverberant signals are mixed with a signal-to-noise ratio (SNR) uniformly sampled from -5 dB to 5 dB. The mixtures are generated by mixing each sample from a different event class. After obtaining the mixed signal, background noise with an SNR of 40 dB is added. All the input audios are resampled to 16kHz. 

In the experiment, a microphone is installed on a rigid ball with a radius of 8$cm$. The microphone is positioned on the equatorial plane of a rigid sphere parallel to the ground. The room impulse response (RIR) is simulated according to \cite{jarrett2012rigid}. The room size is uniformly sampled from $3.0m \times 3.0m \times 2.5m$ to $8.0m \times  8.0m \times 4.0m$. Reverberation Time (RT60) is sampled from $0.2s$ to $0.8s$. The position of the rigid ball and the sound source are guaranteed to be at least 0.8$m$ away from the wall, and the distance between the sound source and the center of the rigid ball is sampled within the range from 0.6$m$ to 2.0$m$. Pitch information is extracted from the single direct-path sound source with Praat \cite{boersma2001speak}. RIRs for training, validation, and testing are 10000, 2000, and 5000. The total number of reverberation samples is 50000, 5000, and 5000, respectively.
\subsection{Experimental details}
For conditional pitch extraction, the frequency of pitch ranges from C1 (32.7 Hz) to B6 (1975.5 Hz) with 20 cents of intervals in the logarithmic scale \cite{wei23b_interspeech}. Cross-entropy is used as the loss function \cite{li2023pgss}. The learning rate and batch size are set to $10^{-4}$ and 32, respectively. The network model, optimized with Adam \cite{kingma2014adam},  is implemented using pytorch\_lightning \cite{falcon2019pytorch}, and distributed data-parallel (DDP) is set to achieve data parallelism on multiple GPUs. To measure the accuracy of conditional pitch estimation, Raw Pitch Accuracy (RPA) is used to achieve the pitch estimation results of frame-by-frame signals \cite{wei23b_interspeech}. Cosine similarity (COSS) is chosen to evaluate the estimation performance for the sequence-level pitch accuracy.

For target sound extraction, the configuration of the network training remains the same as \cite{veluri2023semantic}. Batch size and training epochs are 32 and 80, respectively. The learning rate is initialized as 5 $\times$ $10^{-4}$ while halving the learning rate after 40 epochs. The network is trained with a combination of 90\% SNR and 10\% scale-invariant-signal-to-noise-ratio (SI-SNR) loss. The improvements of SNR and SI-SNR (SNRi, SI-SNRi) are evaluated for the extracted sound.

\section{Results and Discussion}
\subsection{Pitch extraction performances}
\begin{table}[htb]
\centering
\caption{Experimental results on conditional pitch extraction under different model depths and model structures.}
\resizebox{0.47\textwidth}{!}{
\begin{tabular}{c|ccc|ccc}
\hline
\textbf{Depth} & \multicolumn{3}{c|}{\textbf{RPA (\%)}} & \multicolumn{3}{c}{\textbf{COSS}} \\
 & Concat & FiLMAtten & FILM  & Concat & FILMAtten & FILM  \\
\hline
4 & 69.60 & 70.72 & 71.12 & 0.9473 & 0.9484 & 0.9502 \\
5 & 70.81 & 71.89 & 72.17 & 0.9526 & 0.9522 & 0.9534 \\
6 & 71.62 & 73.25 & 73.02 & 0.9520 & 0.9566 & 0.9571 \\
7 & 72.24 & 72.47 & 73.95 & 0.9548 & 0.9550 & 0.9588 \\
8 & 73.73 & 74.18 & 75.16 & 0.9579 & 0.9571 & 0.9607 \\
9 & 72.70 & 75.30 & \textbf{75.52} & 0.9549 & 0.9612 & 0.9625 \\
10 & 73.42 & 74.96 & 75.40 & 0.9553 & 0.9589 & \textbf{0.9626} \\
\hline
\end{tabular}
}
\label{table:method_comparison_f0}
\end{table}

Table \ref{table:method_comparison_f0} shows the performance of conditional pitch extraction under different numbers of TCN layers. To compare the effects of condition inputs, concatenate, named Concat, is used for comparison. Besides, a newly proposed attention-based TCN method (short as FiLMAtten) is introduced for comparison \cite{opochinsky2024single}. The results show that the performance improves as the TCN layer's depth increases. Deeper TCN models can effectively extract frequency characteristics and capture harmonic patterns, leading to more accurate pitch extraction results in reverberant environments. The results also show that the FiLM method is better than the Concat method. Adding the attention mechanism does not bring advantages in pitch extraction.

\begin{figure*}[ht]
    \centering
    \includegraphics[width=0.86\textwidth]{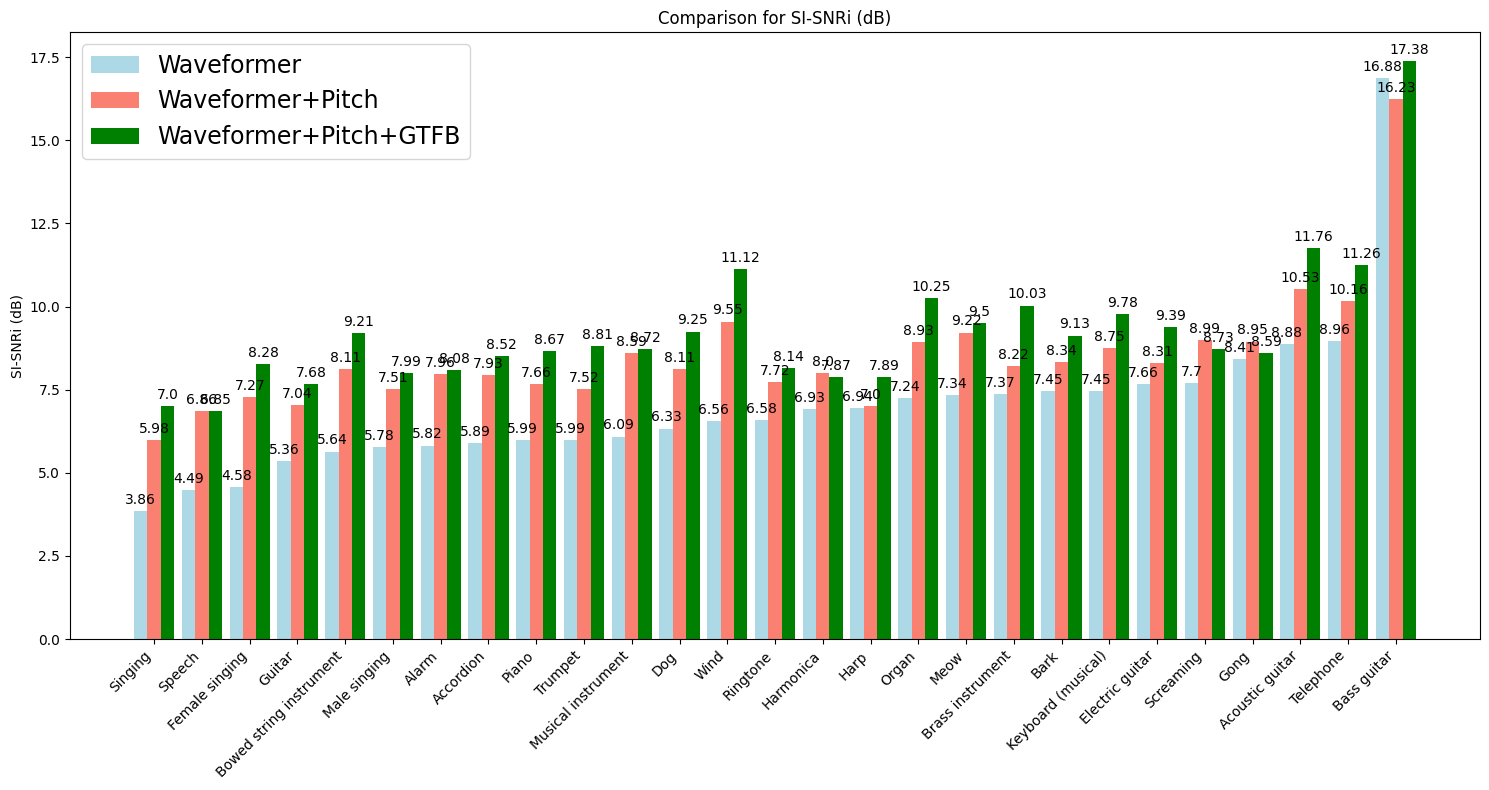}
    \caption{Comparison results for SI-SDRi (dB) with different conditions under reverberant conditions.}
    \label{fig:cmp_sisnr_plot}
\end{figure*}

\subsection{Target sound separation performances}

\begin{table}[htbp]
\centering
\caption{Target sound extraction results under different conditions.}
\resizebox{0.49\textwidth}{!}{
\begin{tabular}{c|c|cc}
\hline
\textbf{Model} & \textbf{Reverb} & \textbf{SNRi(dB)} & \textbf{SiSNRi(dB)} \\
\hline
DPRNN PIT & w/o & — & 12.59 \\
% \hline
DPRNN PIT & w & — & 8.90 \\
\hline
Waveformer & w/o & 6.28 & 9.23 \\
% \hline
Waveformer & w & 5.81 & 7.10 \\
\hline
Waveformer+Pitch & w & 7.07 & 8.91 \\
\hline
\end{tabular}
}
\label{tab:performance_comparison}
\end{table}

\begin{table}[htbp]
\centering
\caption{Ratio between SNR ($\omega_{1}$) and SI-SNR ($\omega_{2}$) under reverberant conditions.}
\resizebox{0.32\textwidth}{!}{
\begin{tabular}{c|c|c|c}
\hline
\textbf{$\omega_{1}$} & \textbf{$\omega_{2}$} & \textbf{SNRi(dB)} & \textbf{SI-SNRi(dB)} \\
\hline
0.5 & 0.5 & 2.15 & 8.89 \\
\hline
0.7 & 0.3 & 2.20 & 7.79 \\
\hline
0.9 & 0.1 & \textbf{7.07} & \textbf{8.91} \\
\hline
\end{tabular}
}
\label{tab:weight_impact_pitchwaveformer}
\end{table}

Table \ref{tab:performance_comparison} compares the performance of target sound extraction. DPRNN \cite{luo2020dual}, trained with permutation invariant training (PIT) \cite{yu2017permutation}, is chosen as a reference method. Under reverberant conditions, the performance of both DPRNN and Waveformer is reduced. Pitch information obtained in the first stage improves the model’s performance under reverberant conditions. Besides, the training ratio for SNR and SI-SNR loss is verified under reverberant conditions, as shown in Table \ref{tab:weight_impact_pitchwaveformer}. The results suggest that the training ratio of the two loss functions is similar to the anechoic sound results \cite{veluri2023real}.

\begin{table}[htbp]
\caption{Comparison results for the learnable GTFB.}
\centering
\resizebox{0.50\textwidth}{!}{
\begin{tabular}{c|c|c|c|c}
\hline
\textbf{Filter Type} & \textbf{Length} & \textbf{Number} & \textbf{SNRi (dB)} & \textbf{SI-SNRi (dB)} \\
\hline
GTFB (l)  & 8 & 512 & 7.45 & 9.37 \\
GTFB (l) & 32 & 512 & \textbf{7.57} & \textbf{9.51} \\
GTFB (l) & 8 & 256 & 6.59 & 8.39 \\
GTFB (l) & 32 & 256 & -2.56 &  7.83\\
\hline
GTFB (f) & 8 & 512 & 7.09 & 8.95\\
GTFB (f) & 32 & 512 & 6.67 & 8.44 \\
\hline
Conv 1D & 8 & 512 & 7.07 & 8.91\\
Conv 1D & 8 & 256 & 6.74 & 8.49 \\
Conv 1D & 32 & 512 & -2.57 & 8.45 \\
Conv 1D & 32 & 256 & -2.56 & 8.28 \\
\hline
\end{tabular}
}
\label{tab:filter_type_performance}
\end{table}

Table \ref{tab:filter_type_performance} shows the results for GTFB, where (l) represents the learnable GTFB and (f) is short for fixed parameters. The results show that learnable GTFB can better utilize pitch information than the 1-D convolution encoder. Different filter lengths and kernel numbers are used to select optimal parameters. Under the optimal parameters, the reverberation performance based on GTFB (SI-SNRi, 9.51 dB) surprisingly exceeds the results without reverberation (9.23 dB). The proposed TSE-PI brings about 2.4 dB SI-SNR improvements compared with the original Waveformer. However, it should be pointed out that the performance under different parameter conditions is quite different, which remains to be further analyzed in subsequent studies.

Figure \ref{fig:cmp_sisnr_plot} compares the SI-SNRi results under optimal parameters for each class. The results show that pitch information provides improvements under most conditions. Using GTFB further improves the performance. This result validates our analysis of the robustness of the auditory system and confirms the effectiveness of the bottom-up process in ASA mechanisms. 

\section{Conclusion}
This paper proposes a novel target sound extraction model with pitch information (TSE-PI). Inspired by the human auditory system, pitch information and Gammatone filterbanks are introduced to improve performance under reverberant conditions. We plan to extend our method to multiple microphones under real-world reverberant scenarios with self-supervised schemes.

\section{Acknowledgement}
This work is supported in part by the Major Program of the National Social Science Fund of China (No. 22\&ZD318), and the High-performance Computing Platform of Peking University.

\newpage

\bibliographystyle{IEEEtran}
\bibliography{mybib}

\begin{thebibliography}{10}
\providecommand{\url}[1]{#1}
\csname url@samestyle\endcsname
\providecommand{\newblock}{\relax}
\providecommand{\bibinfo}[2]{#2}
\providecommand{\BIBentrySTDinterwordspacing}{\spaceskip=0pt\relax}
\providecommand{\BIBentryALTinterwordstretchfactor}{4}
\providecommand{\BIBentryALTinterwordspacing}{\spaceskip=\fontdimen2\font plus
\BIBentryALTinterwordstretchfactor\fontdimen3\font minus
  \fontdimen4\font\relax}
\providecommand{\BIBforeignlanguage}[2]{{%
\expandafter\ifx\csname l@#1\endcsname\relax
\typeout{** WARNING: IEEEtran.bst: No hyphenation pattern has been}%
\typeout{** loaded for the language `#1'. Using the pattern for}%
\typeout{** the default language instead.}%
\else
\language=\csname l@#1\endcsname
\fi
#2}}
\providecommand{\BIBdecl}{\relax}
\BIBdecl

\bibitem{cherry1953cocktail}
C.~Cherry, ``Cocktail party problem,'' \emph{Journal of the Acoustical Society
  of America}, vol.~25, pp. 975--979, 1953.

\bibitem{gfeller2021one}
B.~Gfeller, D.~Roblek, and M.~Tagliasacchi, ``One-shot conditional audio
  filtering of arbitrary sounds,'' in \emph{ICASSP 2021-2021 IEEE International
  Conference on Acoustics, Speech and Signal Processing (ICASSP)}.\hskip 1em
  plus 0.5em minus 0.4em\relax IEEE, 2021, pp. 501--505.

\bibitem{tzinis22_interspeech}
E.~Tzinis, G.~Wichern, A.~S. Subramanian, P.~Smaragdis, and J.~{Le Roux},
  ``{Heterogeneous Target Speech Separation},'' in \emph{Proc. Interspeech
  2022}, 2022, pp. 1796--1800.

\bibitem{delcroix2022soundbeam}
M.~Delcroix, J.~B. V{\'a}zquez, T.~Ochiai, K.~Kinoshita, Y.~Ohishi, and
  S.~Araki, ``Soundbeam: Target sound extraction conditioned on sound-class
  labels and enrollment clues for increased performance and continuous
  learning,'' \emph{IEEE/ACM Transactions on Audio, Speech, and Language
  Processing}, vol.~31, pp. 121--136, 2022.

\bibitem{ohishi2022conceptbeam}
Y.~Ohishi, M.~Delcroix, T.~Ochiai, S.~Araki, D.~Takeuchi, D.~Niizumi,
  A.~Kimura, N.~Harada, and K.~Kashino, ``Conceptbeam: Concept driven target
  speech extraction,'' in \emph{Proceedings of the 30th ACM International
  Conference on Multimedia}, 2022, pp. 4252--4260.

\bibitem{veluri2023real}
B.~Veluri, J.~Chan, M.~Itani, T.~Chen, T.~Yoshioka, and S.~Gollakota,
  ``Real-time target sound extraction,'' in \emph{ICASSP 2023-2023 IEEE
  International Conference on Acoustics, Speech and Signal Processing
  (ICASSP)}.\hskip 1em plus 0.5em minus 0.4em\relax IEEE, 2023, pp. 1--5.

\bibitem{veluri2023semantic}
B.~Veluri, M.~Itani, J.~Chan, T.~Yoshioka, and S.~Gollakota, ``Semantic
  hearing: Programming acoustic scenes with binaural hearables,'' in
  \emph{Proceedings of the 36th Annual ACM Symposium on User Interface Software
  and Technology}, 2023, pp. 1--15.

\bibitem{zmolikova2023neural}
K.~Zmolikova, M.~Delcroix, T.~Ochiai, K.~Kinoshita, J.~{\v{C}}ernock{\`y}, and
  D.~Yu, ``Neural target speech extraction: An overview,'' \emph{IEEE Signal
  Processing Magazine}, vol.~40, no.~3, pp. 8--29, 2023.

\bibitem{hai2024dpm}
J.~Hai, H.~Wang, D.~Yang, K.~Thakkar, N.~Dehak, and M.~Elhilali, ``Dpm-tse: A
  diffusion probabilistic model for target sound extraction,'' in \emph{ICASSP
  2024-2024 IEEE International Conference on Acoustics, Speech and Signal
  Processing (ICASSP)}.\hskip 1em plus 0.5em minus 0.4em\relax IEEE, 2024, pp.
  1196--1200.

\bibitem{kim2024improving}
D.~Kim, M.-S. Baek, Y.~Kim, and J.-H. Chang, ``Improving target sound
  extraction with timestamp knowledge distillation,'' in \emph{ICASSP 2024-2024
  IEEE International Conference on Acoustics, Speech and Signal Processing
  (ICASSP)}.\hskip 1em plus 0.5em minus 0.4em\relax IEEE, 2024, pp. 1396--1400.

\bibitem{chen2022zero}
K.~Chen, X.~Du, B.~Zhu, Z.~Ma, T.~Berg-Kirkpatrick, and S.~Dubnov, ``Zero-shot
  audio source separation through query-based learning from weakly-labeled
  data,'' in \emph{Proceedings of the AAAI Conference on Artificial
  Intelligence}, vol.~36, no.~4, 2022, pp. 4441--4449.

\bibitem{liu22w_interspeech}
X.~Liu, H.~Liu, Q.~Kong, X.~Mei, J.~Zhao, Q.~Huang, M.~D. Plumbley, and
  W.~Wang, ``{Separate What You Describe: Language-Queried Audio Source
  Separation},'' in \emph{Proc. Interspeech 2022}, 2022, pp. 1801--1805.

\bibitem{li2023target}
C.~Li, Y.~Qian, Z.~Chen, D.~Wang, T.~Yoshioka, S.~Liu, Y.~Qian, and M.~Zeng,
  ``Target sound extraction with variable cross-modality clues,'' in
  \emph{ICASSP 2023-2023 IEEE International Conference on Acoustics, Speech and
  Signal Processing (ICASSP)}.\hskip 1em plus 0.5em minus 0.4em\relax IEEE,
  2023, pp. 1--5.

\bibitem{ye2024lavss}
Y.~Ye, W.~Yang, and Y.~Tian, ``Lavss: Location-guided audio-visual spatial
  audio separation,'' in \emph{Proceedings of the IEEE/CVF Winter Conference on
  Applications of Computer Vision}, 2024, pp. 5508--5519.

\bibitem{veluri2024look}
B.~Veluri, M.~Itani, T.~Chen, T.~Yoshioka, and S.~Gollakota, ``Look once to
  hear: Target speech hearing with noisy examples,'' in \emph{Proceedings of
  the CHI Conference on Human Factors in Computing Systems}, 2024, pp. 1--16.

\bibitem{lee2023deft}
D.~Lee and J.-W. Choi, ``Deft-an: Dense frequency-time attentive network for
  multichannel speech enhancement,'' \emph{IEEE Signal Processing Letters},
  vol.~30, pp. 155--159, 2023.

\bibitem{choi2023target}
D.~Choi and J.-W. Choi, ``Target sound extraction on reverberant mixture,''
  \emph{The Journal of the Acoustical Society of America}, vol. 154, no.
  4\_supplement, pp. A270--A271, 2023.

\bibitem{bregman1994auditory}
A.~S. Bregman, \emph{Auditory scene analysis}.\hskip 1em plus 0.5em minus
  0.4em\relax Citeseer, 1994, vol. 198.

\bibitem{wang2006computational}
D.~Wang and G.~J. Brown, \emph{Computational auditory scene analysis:
  Principles, algorithms, and applications}.\hskip 1em plus 0.5em minus
  0.4em\relax Wiley-IEEE press, 2006.

\bibitem{shamma2011temporal}
S.~A. Shamma, M.~Elhilali, and C.~Micheyl, ``Temporal coherence and attention
  in auditory scene analysis,'' \emph{Trends in neurosciences}, vol.~34, no.~3,
  pp. 114--123, 2011.

\bibitem{lopez2001human}
E.~A. Lopez-Poveda and R.~Meddis, ``A human nonlinear cochlear filterbank,''
  \emph{The Journal of the Acoustical Society of America}, vol. 110, no.~6, pp.
  3107--3118, 2001.

\bibitem{kim2018crepe}
J.~W. Kim, J.~Salamon, P.~Li, and J.~P. Bello, ``Crepe: A convolutional
  representation for pitch estimation,'' in \emph{2018 IEEE International
  Conference on Acoustics, Speech and Signal Processing (ICASSP)}.\hskip 1em
  plus 0.5em minus 0.4em\relax IEEE, 2018, pp. 161--165.

\bibitem{li2023pgss}
X.~Li, Y.~Wang, Y.~Sun, X.~Wu, and J.~Chen, ``Pgss: pitch-guided speech
  separation,'' in \emph{Proceedings of the AAAI Conference on Artificial
  Intelligence}, vol.~37, no.~11, 2023, pp. 13\,130--13\,138.

\bibitem{singh2021deepf0}
S.~Singh, R.~Wang, and Y.~Qiu, ``Deepf0: End-to-end fundamental frequency
  estimation for music and speech signals,'' in \emph{ICASSP 2021-2021 IEEE
  International Conference on Acoustics, Speech and Signal Processing
  (ICASSP)}.\hskip 1em plus 0.5em minus 0.4em\relax IEEE, 2021, pp. 61--65.

\bibitem{wei23b_interspeech}
H.~Wei, X.~Cao, T.~Dan, and Y.~Chen, ``{RMVPE: A Robust Model for Vocal Pitch
  Estimation in Polyphonic Music},'' in \emph{Proc. INTERSPEECH 2023}, 2023,
  pp. 5421--5425.

\bibitem{perez2018film}
E.~Perez, F.~Strub, H.~De~Vries, V.~Dumoulin, and A.~Courville, ``Film: Visual
  reasoning with a general conditioning layer,'' in \emph{Proceedings of the
  AAAI conference on artificial intelligence}, vol.~32, no.~1, 2018.

\bibitem{luo2019conv}
Y.~Luo and N.~Mesgarani, ``Conv-tasnet: Surpassing ideal time--frequency
  magnitude masking for speech separation,'' \emph{IEEE/ACM transactions on
  audio, speech, and language processing}, vol.~27, no.~8, pp. 1256--1266,
  2019.

\bibitem{li2021auditory}
H.~Li, K.~Chen, and B.~U. Seeber, ``Auditory filterbanks benefit universal
  sound source separation,'' in \emph{ICASSP 2021-2021 IEEE International
  Conference on Acoustics, Speech and Signal Processing (ICASSP)}.\hskip 1em
  plus 0.5em minus 0.4em\relax IEEE, 2021, pp. 181--185.

\bibitem{fonseca2021fsd50k}
E.~Fonseca, X.~Favory, J.~Pons, F.~Font, and X.~Serra, ``Fsd50k: an open
  dataset of human-labeled sound events,'' \emph{IEEE/ACM Transactions on
  Audio, Speech, and Language Processing}, vol.~30, pp. 829--852, 2021.

\bibitem{jarrett2012rigid}
D.~P. Jarrett, E.~A. Habets, M.~R. Thomas, and P.~A. Naylor, ``Rigid sphere
  room impulse response simulation: Algorithm and applications,'' \emph{The
  Journal of the Acoustical Society of America}, vol. 132, no.~3, pp.
  1462--1472, 2012.

\bibitem{boersma2001speak}
P.~Boersma and V.~Van~Heuven, ``Speak and unspeak with praat,'' \emph{Glot
  International}, vol.~5, no. 9/10, pp. 341--347, 2001.

\bibitem{kingma2014adam}
D.~P. Kingma and J.~Ba, ``Adam: A method for stochastic optimization,''
  \emph{arXiv preprint arXiv:1412.6980}, 2014.

\bibitem{falcon2019pytorch}
W.~A. Falcon, ``Pytorch lightning,'' \emph{GitHub}, vol.~3, 2019.

\bibitem{opochinsky2024single}
R.~Opochinsky, M.~Moradi, and S.~Gannot, ``Single-microphone speaker separation
  and voice activity detection in noisy and reverberant environments,''
  \emph{arXiv preprint arXiv:2401.03448}, 2024.

\bibitem{luo2020dual}
Y.~Luo, Z.~Chen, and T.~Yoshioka, ``Dual-path rnn: efficient long sequence
  modeling for time-domain single-channel speech separation,'' in \emph{ICASSP
  2020-2020 IEEE International Conference on Acoustics, Speech and Signal
  Processing (ICASSP)}.\hskip 1em plus 0.5em minus 0.4em\relax IEEE, 2020, pp.
  46--50.

\bibitem{yu2017permutation}
D.~Yu, M.~Kolb{\ae}k, Z.-H. Tan, and J.~Jensen, ``Permutation invariant
  training of deep models for speaker-independent multi-talker speech
  separation,'' in \emph{2017 IEEE International Conference on Acoustics,
  Speech and Signal Processing (ICASSP)}.\hskip 1em plus 0.5em minus
  0.4em\relax IEEE, 2017, pp. 241--245.

\end{thebibliography}

\end{document}